# Modular Photobioreactor Façade Systems for Sustainable Architecture - A case study

Xiujin Liu [1]

[1]University of Michigan – Ann Arbor, Ann Arbor, The United States

jeanliu@umich.edu

**Abstract.** This paper proposes an innovative solution to the growing issue of greenhouse gas emissions: a closed photobioreactor (PBR) façade system to mitigate greenhouse gas (GHG) concentrations. With digital fabrication technology, this study explores the transition from traditional, single function building facades to multifunctional, integrated building systems. It introduces a photobioreactor façade system to mitigate greenhouse gas (GHG) concentrations while addressing the challenge of large-scale prefabricated components transportation. This research introduces a novel approach by designing the façade system as modular, user-friendly and transportation-friendly bricks, enabling the creation of a user-customized and self-assembled photobioreactor system. The single module in the system is proposed to be "neutralization bricks", which embedded with algae and equipped with an air circulation system, facilitating the photobioreactor's functionality. A connection system between modules allows for easy assembly by users, while a limited variety of brick styles ensures modularity in manufacturing without sacrificing customization and diversity. The system is also equipped with an advanced microalgae status detection algorithm, which allows users to monitor the condition of the microalgae using monocular camera. This functionality ensures timely alerts and notifications for users to replace the algae, thereby optimizing the operational efficiency and sustainability of the algae cultivation process.

## 1 Introduction

Global warming has been a catalyst for significant environmental changes and impacts on human life. NASA observations indicate an approximate 1°C (1.8°F) increase in Earth's average global temperature relative to pre-industrial levels.

This warming trend is occurring at an unprecedented rate of over 0.2°C (0.36°F) per decade.
Fortunately, strategies to mitigate these emissions through increased energy efficiency and the adoption of renewable energy sources, such as solar and wind, are being explored. *Because of the high growth rates and CO2 bio fixation potentials, the large-scale cultivation of microalgae has been proposed as potentially a sustainable solution to the capture and recycle of CO2 emissions* (Subhadra 2011, Wilson et al. 2021). This paper explores the potential combination of microalgae and building component to mitigate greenhouse gas concentrations. we introduce a case study with innovative brick-based microalgae photobioreactor (PBR) façade system. In this conceptual project that simulates structural and functional behavior, by integrating essential PBR functionalities within small-scale bricks, we achieve the following innovations:
**1.** Achieves sustainable building components, by using microalgae to mitigate CO2 emissions as well as naturally regulate building temperatures
**2.** Challenges conventional construction methods and highlights the system's flexibility and ease of installation, so that reduces reliance on labor-intensive manufacturing processes.
**3.** Provides an innovative modularity PBR Façade System equipped with advanced monitoring technology

## 2 Related Works

Photobioreactors (PBRs) are advanced systems that use sunlight to grow photoautotrophic organisms like microalgae. Their application on building façades shows great potential. The BIQ building in Hamburg uses algae façades for shading and $CO_2$ conversion (Schiller, 2014). Kim (2013) developed a flexible algae façade that promotes wellness and produces biofuel with a closed-loop system. Lee et al. (2014) created the Pleura Pod, which filters air through algae pods to convert $CO_2$ into oxygen. Arora et al. (2024) explore how integrating microalgal PBRs in buildings can reduce energy use and emissions while supporting sustainable urban living.
Modular architecture offers efficient, cost-effective construction using prefabricated units, with benefits like faster installation, quality control, and reusability (Lawson et al., 2012). It also eases transport and maintenance. Shroff and Aditya (2022) explored modular strategies for flexible, sustainable urban design, while Park (2017) proposed an affordable modular home in Korea. Shin and Park (2025) addressed facade design diversity in modular buildings, suggesting creative strategies to improve aesthetics and industry acceptance.
Despite of these advancements, gaps remain unaddressed:

**1.** Existing research often focuses on theoretical and small-scale experimental validations, neglecting practical challenges of large-scale implementation. This complexity limits their applicability and effectiveness in real-world, urban environments.
**2.** Moreover, there is often a lack of balance between customization and manufacturing efficiency in modular PBR designs, especially when considering transportations. Most modular systems offer limited customization options, restricting their integration into diverse architectural contexts and reducing their aesthetic and functional appeal.
**3.** Additionally, current PBR systems often lack advanced, real-time monitoring solutions for algae health and system performance using accessible equipment, existing systems do not provide user-friendly, real-time feedback mechanisms, resulting in suboptimal operation and reduced sustainability.

## 3 Methodology

### 3.1 Module Geometry Formulation

Based on modularity design, our first goal is to determine geometric forms that optimize the structural integrity and versatility required for an innovative modular photobioreactor (PBR) façade system. Essential to this pursuit was the ability for the chosen geometries to support vertical and diagonal stacking to facilitate the construction of both straight and curved walls, respectively. These geometries needed to allow for a diverse range of multiple distinct styles that could be combined in a randomized manner while ensuring that the interfaces remained straight and parallel to simplify disassembly processes.
After a comprehensive analysis of regular polyhedrons, examining shapes from tetrahedrons with four surfaces to dodecahedrons with twelve, we identified hexahedrons and octahedrons as the most promising candidates due to their structural characteristics conducive to assembly, such as surfaces with parallel counterparts and an equal number of sides allowing for a reciprocal exchange between vertices and surface counts. Furthermore, it was observed that connecting the center points across the faces of a regular hexahedron forms a regular octahedron, and vice versa, with both structures featuring rectangular middle sections, underscoring their suitability for the envisioned PBR façade system. Figure 1. shows the relationship between hexahedron and octahedron.
In addition, to delve deeper into the assembly potential of hexahedrons and octahedrons, we structured our investigation around two primary variables: angle differences and side lengths (Figure 2.). Finally, Figure 3. Shows the three selected cells.

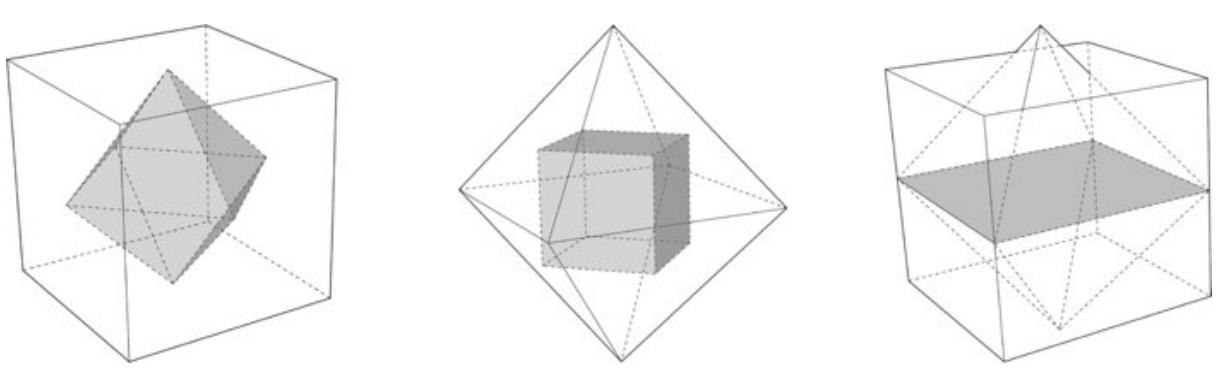

**Figure 1.** Hexahedron and Octahedron Relationship

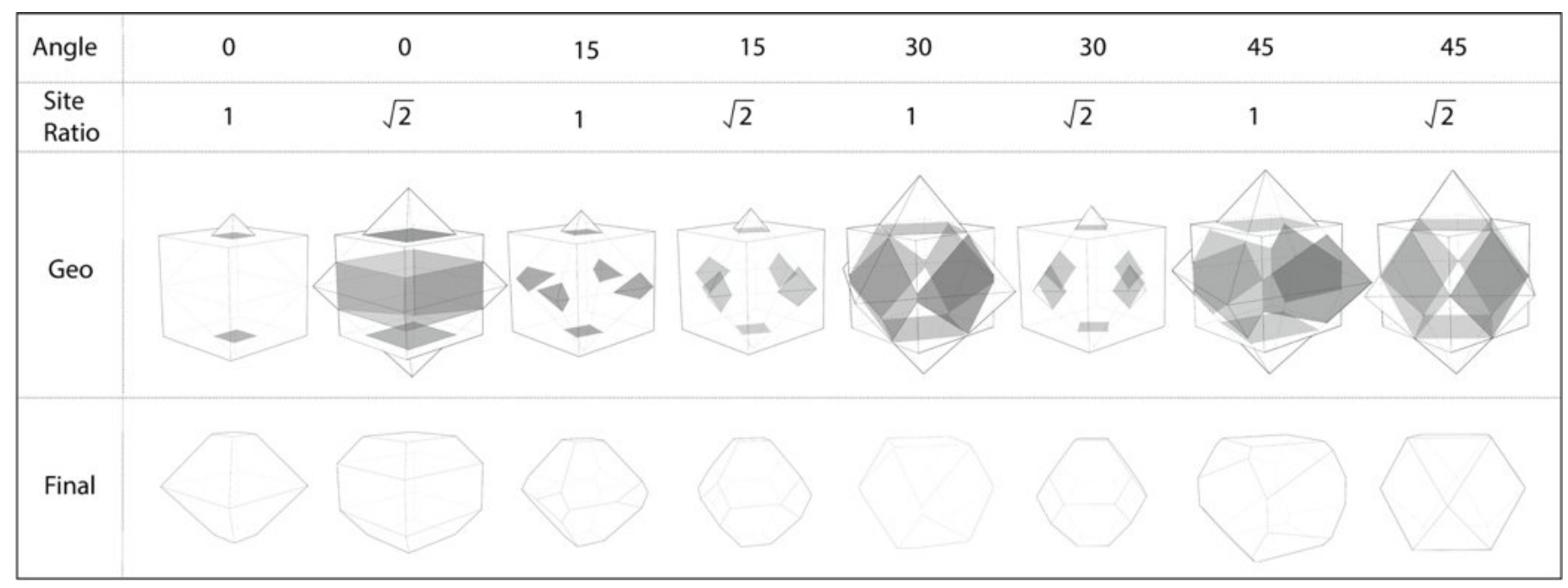

**Figure 2.** Different Combination of Hexahedron and Octahedron Exploration -- Angle Difference: Four groups were established with angle differences set at 0°, 15°, 30°, and 45°. This segmentation allowed us to examine the impact of angular variation on the creation of axisymmetric and centrosymmetric geometries, which are crucial for maintaining a balance between structural assembly rules and customization flexibility; Side Length: We explored the geometry by dividing the experiments into two groups based on side lengths, one maintaining identical lengths for all sides and the other aligning the length of the octahedron's sides with the diagonal length of a regular hexahedron.

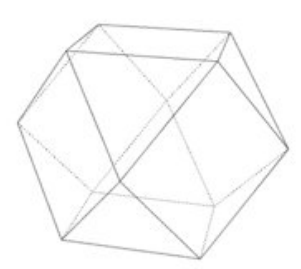
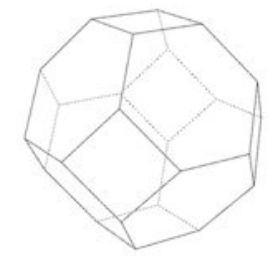
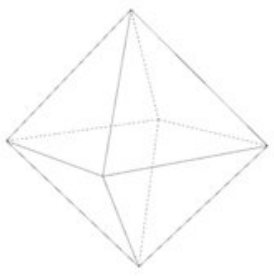

**Figure 3.** Selected Cell Geometry

### 3.2 Operational Mechanism of Individual module

Our study focuses on the development and implementation of microalgae-based closed photobioreactor (PBR) façade systems, which require specific components, including air, fresh algae, nutrient-rich water, an air pump, and piping. The fundamental operating principle involves using an air pump to introduce carbon dioxide-rich "dirty air" into the water. The algae then convert the carbon dioxide into oxygen, which is subsequently released.

In our study, the entire photobioreactor (PBR) façade system is composed of multiple individual cells, each with a consistent operational mechanism that ensures seamless integration and functionality when connected to other cells. Each cell operates by utilizing an air pump to introduce carbon dioxide-rich air into the water, where the algae convert the carbon dioxide into oxygen through photosynthesis, with the oxygen being subsequently released. Figure 4. presents a longitudinal section of a single cell, illustrating the intricate details of its operating mechanism. This includes the integration of piping and the air circulation system that ensures efficient gas exchange and optimal algae growth conditions.

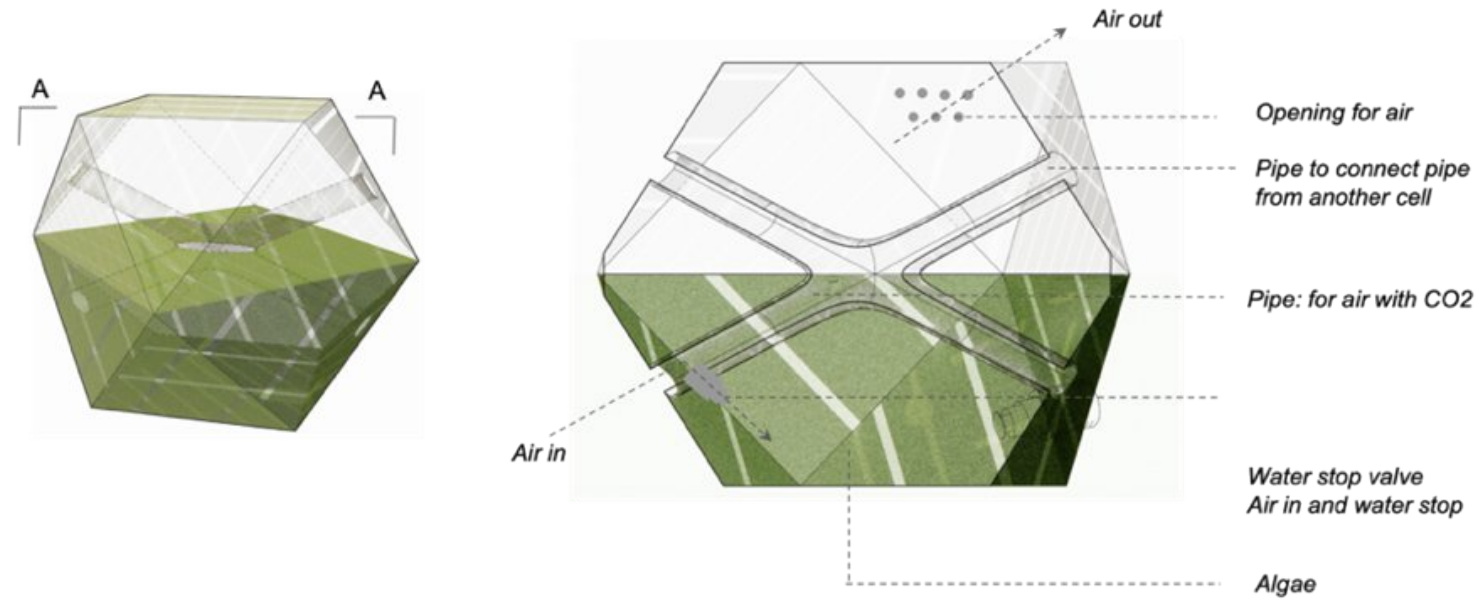

**Figure 4.** Longitudinal Section of a Single Cell

### 3.3 PBR Façade and Piping System Prototype

For our system prototype, the microalgae PBR requires a container for holding microalgae, a device for pumping gas, and a pipeline for transporting gas. Aiming to address challenges related to transportation, assembly, and customization, our methodology integrates the benefits of both holistic and combined system designs. This involves incorporating piping within separate containers and exploring a unified piping system. The goal is to devise a combined system that maintains microalgae separation while minimizing the need for external pumps and pipes. This innovative approach is expected to streamline the assembly process, enhance transportability, and allow for greater customization of the PBR façade system, ultimately contributing to more efficient and adaptable photobioreactor designs. Figure 5. illustrates the prototyped algae based closed PBR system. In this design, the entire façade is divided into cells, with each cell acting as a container for algae. Pipes are inserted inside each cell, allowing for the connection of adjacent cells to form an integrated pipeline system. This system enables the connection of cells to each other, allowing users to transport air to all parts of the façade through only one air pump.

In addition, to enhance the variability, interactivity, and convenience of the system, we have developed a centralized pipeline prototype, shown in Figure 6. This prototype demonstrates the potential configurations and possibilities of the pipeline system, as illustrated in Figure 7.

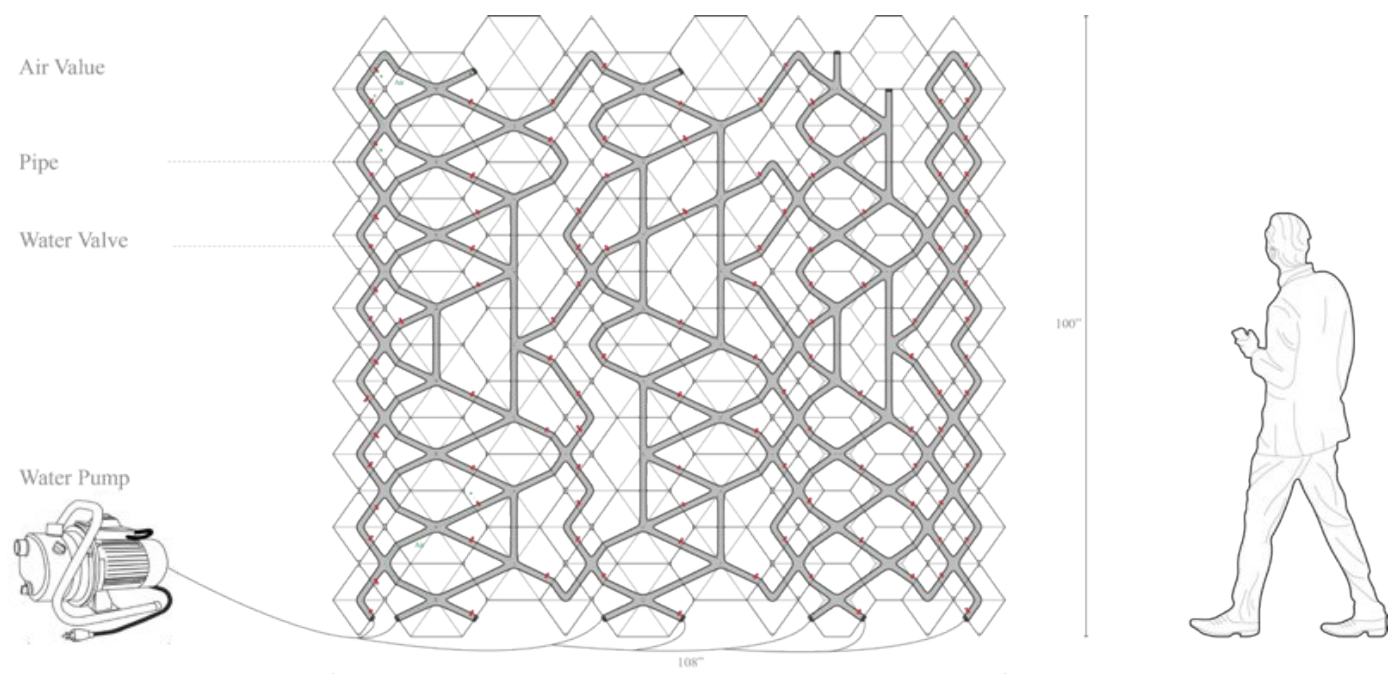

**Figure 5.** Assembly Example of the PBR Façade System

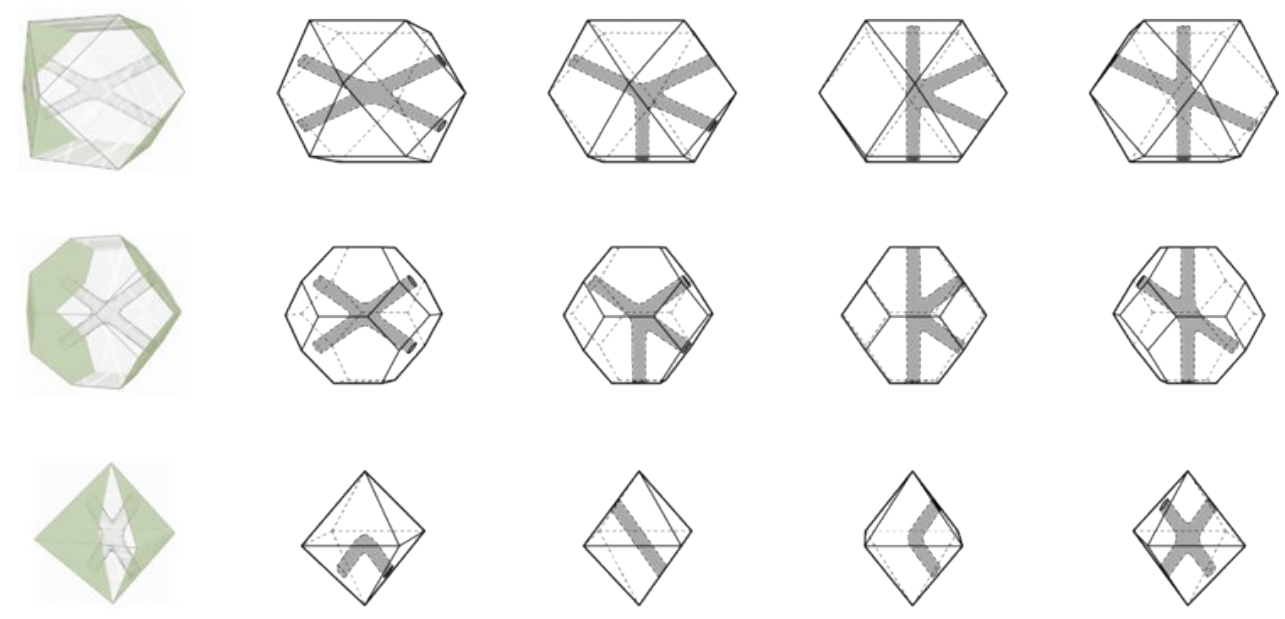

**Figure 6.** Centralized Pipeline Prototype

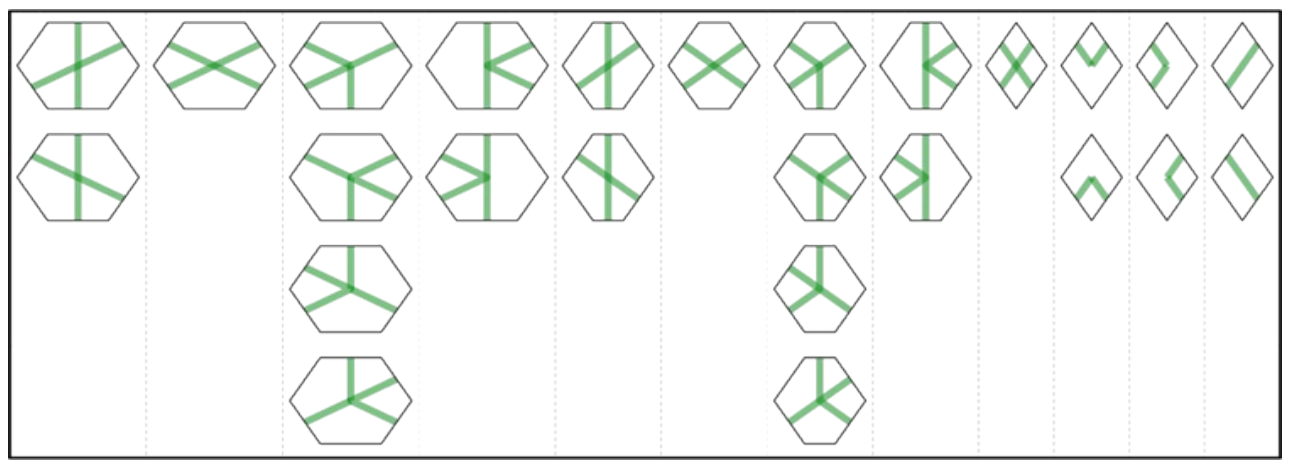

**Figure 7.** Potential Configuration Positions by Rotating Cells

### 3.4 Connection System

Given the modular nature of the entire system, our goal is to enable users to maintain individual modules without compromising the overall system structure. We propose a connection method utilizing magnets, allowing users to remove and maintain single modules seamlessly. Figure 8. illustrates the comparison between the normal and tangential forces exerted by the magnets, highlighting that the tangential force is significantly smaller than the normal force. This characteristic is advantageous as it maintains the stability of the system structure by securely connecting the modules in the normal direction.

Concurrently, it allows users to slide the modules in the tangential direction for easy removal and maintenance. Figure 9 depicts the application of magnets on the modules, along with a force analysis and the removal process. As shown in the diagram, the magnetic connection system allows users to easily slide individual modules forward and remove them without affecting the balance of the remaining façade. Once removed, the old microalgae can be discarded and replaced with fresh algae and nutrient solution. The refreshed module can then be reinserted into the original system, completing the algae replacement process efficiently and conveniently. This innovative magnetic connection mechanism ensures both structural integrity and ease of maintenance, enhancing the practicality and user-friendliness of the modular PBR façade system.

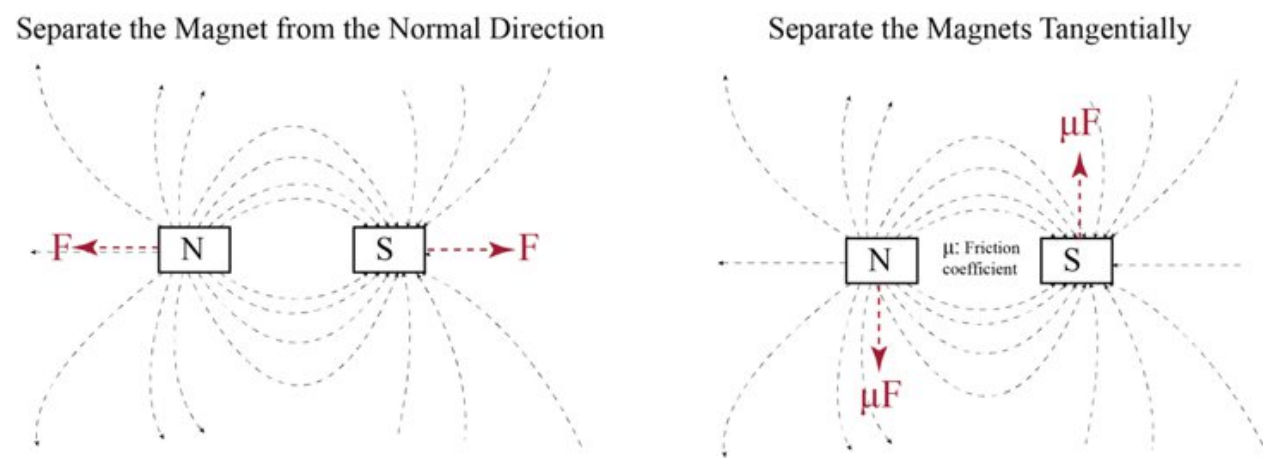

**Figure 8.** Force Comparison

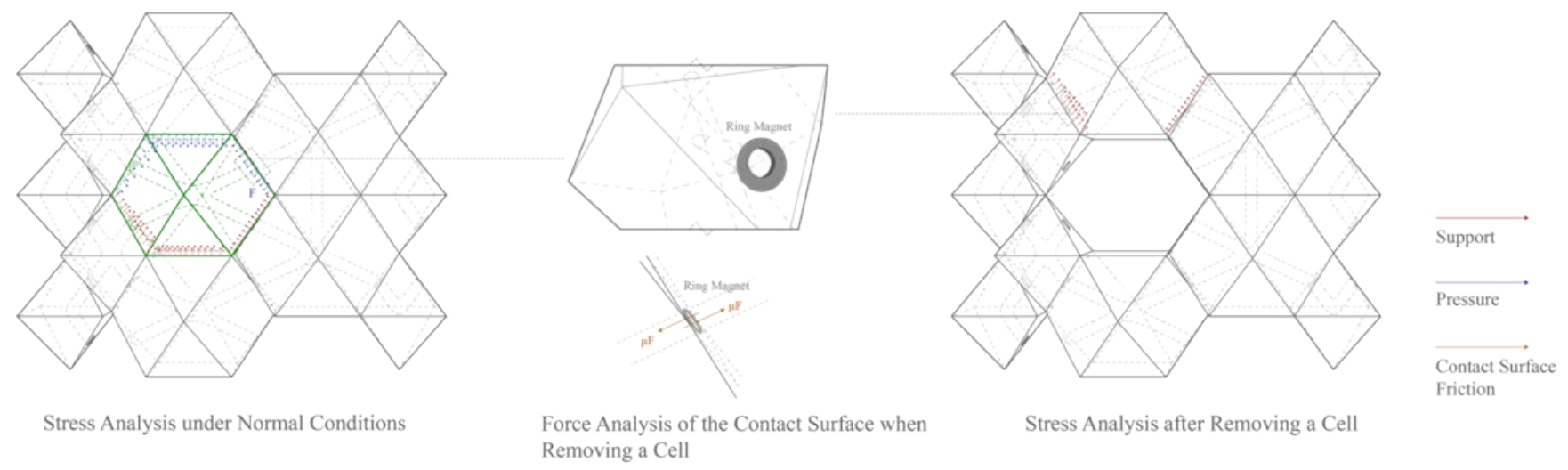

**Figure 9.** Application of Magnets on Modules -- indicates removal process during replacement and force analysis

### 3.5 Microalgae Status Detection Algorithm

In this section, we investigate the algorithm for detecting the status of microalgae by analyzing its RGB color content, because when the microalgae ages, its color shifts from green to red. A monocular RGB camera is used to capture and analyze the microalgae status. To mitigate the influence of varying lighting conditions on color information, a control experiment is established. A pure green label (R: 0, G: 255, B: 0) is used as a control. Under different lighting conditions, the microalgae color is compared to this control color to obtain a regression curve of microalgae growth, which is then used to estimate the status of the microalgae.

For the specific experiment, we first input a series of 2667x2000 RGB images, with each image using its capture time as the ground truth for its state. Figure 10. demonstrates the input image used for the control experiment and our method for color extraction. The left side of the image shows the actual microalgae status recorded by the monocular RGB camera, while the right side shows the color of the green label used for control under the same lighting conditions, position, and camera settings. We employ the 9-grid point method, where spots are taken from nine-grid positions of the image to calculate the RGB values. For each spot, we implement a Gaussian cluster (size 100 and covariance 80) and extract the RGB value of each point in the Gaussian cluster, finally taking the mean value of the Gaussian cluster as the RGB value for this spot. These values are then weighted according to their positions (center point weight is w0 = 0.2, and others are w1 = 0.1) to reduce the bias of any single position and increase the robustness of the entire process.

After obtaining the samples, we estimate the similarity measurement between the test sample and the control sample based on various similarity measurements and then fit the linear, quadratic, and cubic equations to these measurements. Figure 11. shows the regression curve of microalgae state detection we obtained based on different distance functions and fitting functions.

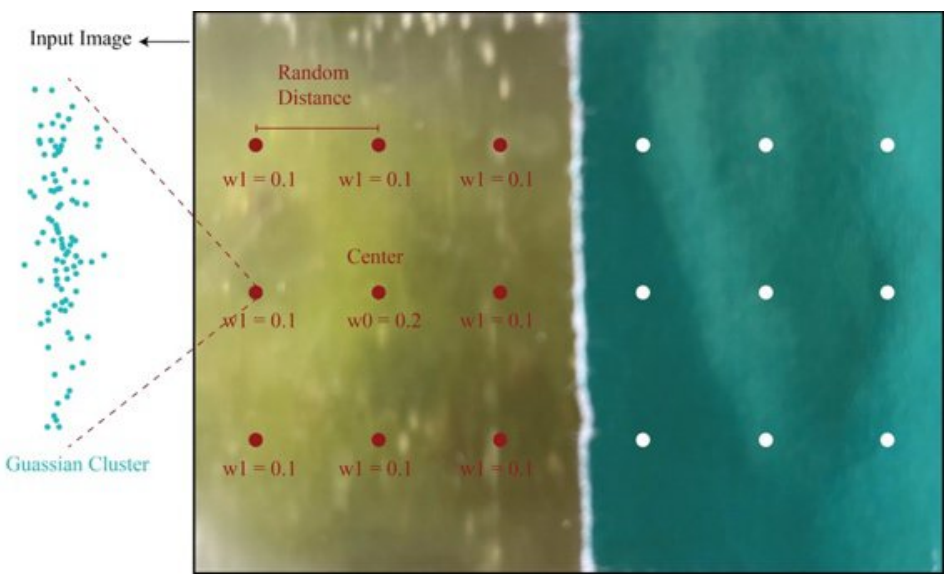

**Figure 10.** Application of Magnets on Modules

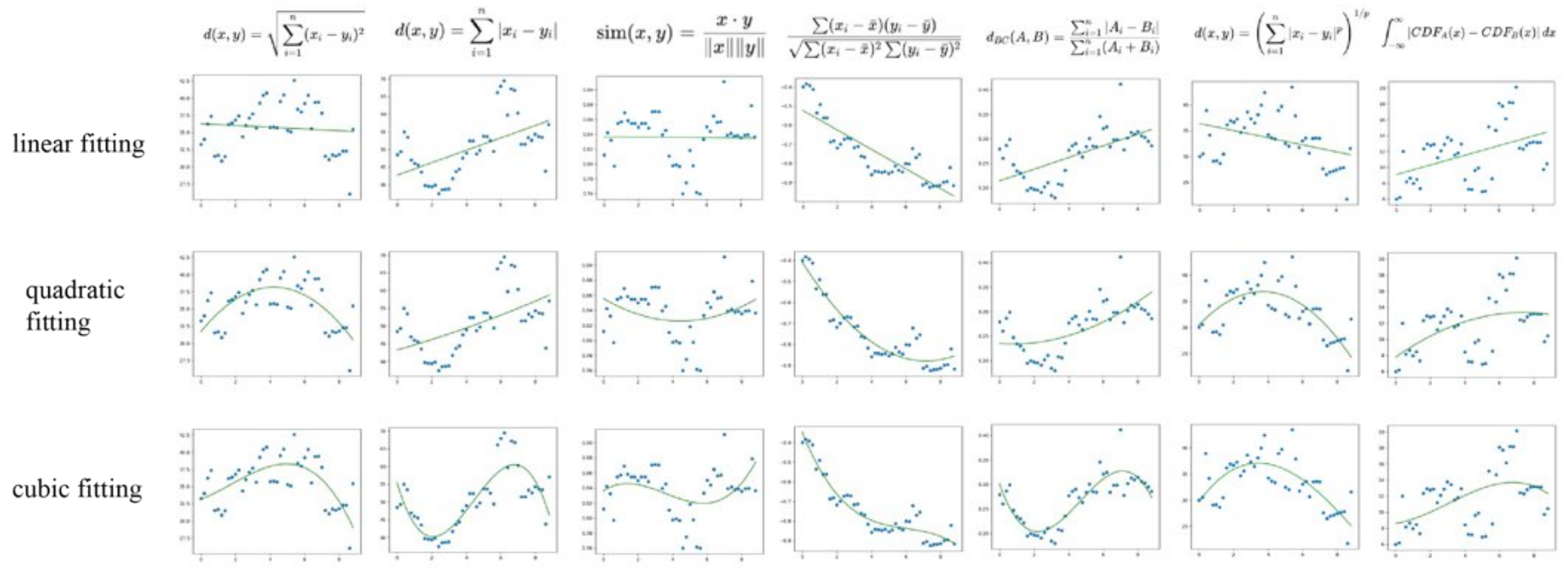

**Figure 11.** Application of Magnets on Modules

## 4 Results and Discussion

The construction and deployment of the PBR façade system were successfully demonstrated using 28 bricks produced with Creality Ender-3 3D printers. This prototype explicitly showcased three types of cells and four types of pipes, forming a 3 ft by 5 ft PBR façade system. The production of each brick took an average of 15 hours and used less than 0.5 kg of PLA, highlighting the process's efficiency and cost-effectiveness.
The innovative design of these cells, featuring embedded piping and modular connectivity, addresses practical challenges such as transportability, assembly, and customization. The modularity of the design allows users to connect cells easily and transport air throughout the façade using an air pump, ensuring efficient operation and adaptability to various architectural contexts. This approach enhances the operational efficiency of the PBR façade system while contributing to environmental sustainability and user-friendliness. Figures 12. visually illustrates the printed cells and the assembled wall, demonstrating the practical implementation of the design.
The microalgae status detection algorithm showed promising results, demonstrating good fitting performance for quadratic and cubic functions, as indicated by the Pearson correlation coefficient. This algorithm, which employs advanced RGB content analysis and the 9-grid point method, enables accurate real-time monitoring of the microalgae's health, ensuring optimal operation of the PBR system. Figure 13 presents the fitting results of the detection algorithm, where negative values correlate positively in absolute terms with the number of days, demonstrating its potential for effective monitoring. It also indicates that the optimal algae replacement cycle is six days

Despite these achievements, several challenges remain. The transition from lab-scale prototypes to industry-scale implementations requires addressing logistical issues related to industry production and on-site assembly. While the modular design offers customization, balancing this with manufacturing efficiency is crucial to ensure the system's economic viability. Evaluating the state of microalgae solely through the regression curve of color and time is a simplistic method with inherent limitations. And additionally, the connection mechanisms between modules, such as the use of magnets, need further optimization to enhance stability and ease of assembly. Future research should focus on refining these connection mechanisms and exploring alternative materials and production methods to improve cost, durability, and environmental impact. Investigating the geometry of the cells and pipes for optimal light and air distribution can also yield improvements in system performance. Moreover, expanding the functionality of the detection algorithm to include more environmental variables and analyze the status of microalgae from multiple dimensions—such as oxygen production, concentration, and microalgae activity—could provide a more comprehensive monitoring solution. To achieve this, the modular design could incorporate additional sensors and analytical techniques, including dissolved oxygen sensors to measure oxygen output, chlorophyll fluorescence sensors to assess photosynthetic efficiency, optical density sensors to estimate microalgae concentration, and pH and $CO_2$ sensors to monitor metabolic activity. Advanced tools like multispectral cameras or imaging flow cytometry could also be used to capture detailed physiological changes and cell viability, while temperature and light sensors would track environmental factors affecting growth. Integrating these technologies would enable a more robust and multi-faceted analysis of microalgae health and performance.

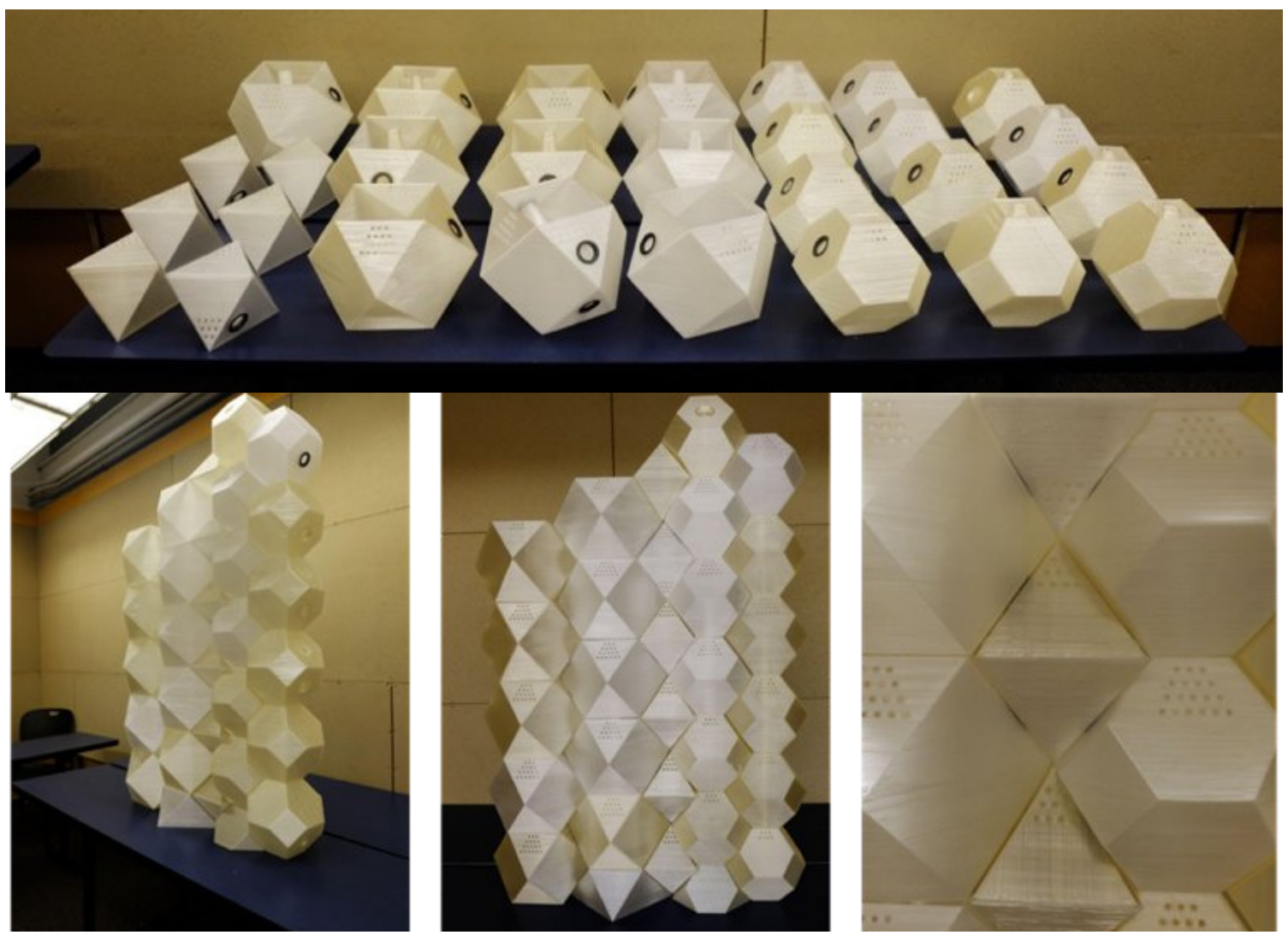

**Figure 12.** Printed Module and Assembled Wall

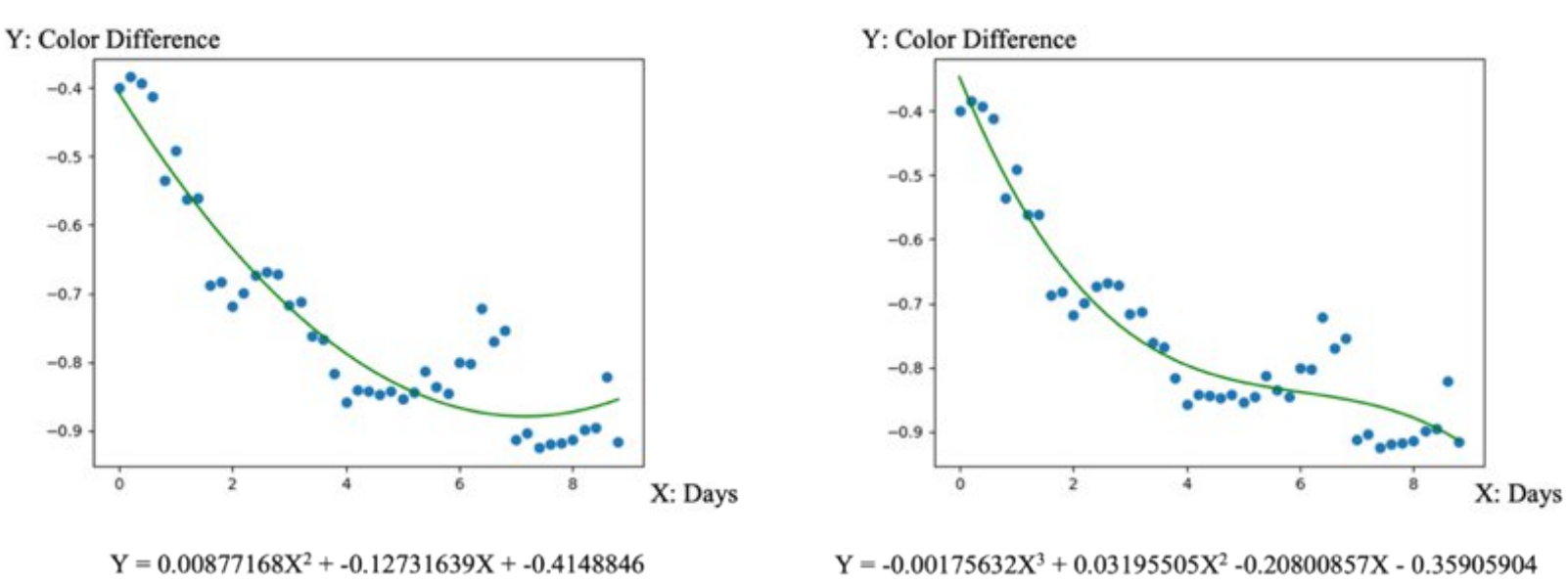

**Figure 13.** Microalgae Status Detection Fitting Results

## 5 Conclusion

This study introduces an innovative closed photobioreactor (PBR) façade system designed to address the increasing issue of greenhouse gas (GHG) emissions due to industrialization and urbanization. By leveraging advancements in digital fabrication, this research transitions traditional, single function building facades into multi-functional, integrated systems that actively reduce GHG concentrations.
The system incorporates "neutralization bricks" embedded with algae and equipped with air circulation systems to facilitate the photobioreactor functionality. A significant aspect of this research is the modular design of the façade, which enables user-friendly assembly and transportation of prefabricated components. This modularity addresses the logistical challenges associated with large-scale implementation, allowing for easy maintenance and replacement of individual modules without compromising the overall structure. The investigation into the geometry of the modules ensures that the design supports both manufacturing efficiency and user customization. By offering a limited variety of brick styles, the system maintains modularity in production while allowing for diverse and aesthetically pleasing architectural configurations. The connection system between modules further enhances ease of assembly and structural stability, enabling users to remove and replace individual modules as needed. An advanced microalgae status detection algorithm is integrated into the system to optimize the cultivation process. Utilizing a monocular RGB camera and sophisticated color analysis technique, this algorithm provides real-time monitoring of the algae's condition. This feature ensures timely alerts for algae replacement, thereby enhancing the operational efficiency and sustainability of the PBR façade.
In conclusion, the proposed closed PBR façade system represents a significant advancement in sustainable building technologies. It offers a practical solution for reducing urban GHG emissions while promoting a user-centric approach to installation and maintenance. The combination of modular design, efficient geometry, and advanced monitoring capabilities underscores the potential of integrating biological processes into architectural designs, paving the way for future innovations in eco-friendly building technologies.

## Acknowledgement

I would like to express my sincere gratitude to my supervisor, Professor Mania Aghaei Meibodi, for her invaluable guidance, support, and encouragement throughout the course of this research. And my teammates, Yuxin Lin and Zhijuan Liu, for their generous help and collaboration.